\documentclass[12pt,letterpaper,a4paper]{article}
\usepackage[includeheadfoot,
            marginratio={1:1,2:3},
            width=412pt,
            height=688pt,]{geometry}
\usepackage{amsmath}
\usepackage{amsfonts}
\usepackage{amssymb}
\usepackage{graphicx}
\usepackage{empheq}
\usepackage{subfigure}
\usepackage{color}
\usepackage{hyperref}
\usepackage{enumerate}

\usepackage{float}
\restylefloat{table}
\restylefloat{figure}

\newcommand{\nc}{\newcommand}
\nc{\beq}{\begin{equation}}
\nc{\eeq}{\end{equation}}
\nc{\bea}{\begin{eqnarray}}
\nc{\eea}{\end{eqnarray}}

\newdimen\csize\csize=1.5ex
\def\young#1{\tiny\vcenter{\hbox{\vrule\vtop{\hrule
  \offinterlineskip\halign{&\vbox
  {\hbox to\csize {\strut\hss##\hss\vrule}\hrule}\cr#1 \crcr}}}}}

\begin{document}


\vspace{1.5cm}
\begin{center}
{\LARGE
Fractional-chaotic inflation in the lights of PLANCK and BICEP2}
\vspace{0.4cm}

\end{center}

\vspace{0.35cm}
\begin{center}
 Xin Gao$^{\dag,\ddag}$\footnote{Email: xingao@vt.edu}, Tianjun Li$^{\dag, \flat}$\footnote{Email: tli@itp.ac.cn}  and Pramod Shukla$^\sharp$\footnote{Email: pkshukla@to.infn.it}
\end{center}

\vspace{0.1cm}
\begin{center}
{\it

$^{\dag}$ State Key Laboratory of Theoretical Physics
and Kavli Institute for Theoretical Physics China (KITPC),
      Institute of Theoretical Physics, Chinese Academy of Sciences,
Beijing 100190, P. R. China \\
\vspace{0.4cm}
$^{\ddag}$ Department of Physics, Robeson Hall, 0435,  Virginia Tech,  \\
850 West Campus Drive, Blacksburg, VA 24061, USA\\
\vspace{0.4cm}
$^{\flat}$ School of Physical Electronics,
University of Electronic Science and Technology of China,
Chengdu 610054, P. R. China \\
\vspace{0.4cm}
$^{\sharp}$ Universit\'a di Torino, Dipartimento di Fisica and I.N.F.N. - sezione di Torino \\
Via P. Giuria 1, I-10125 Torino, Italy
}

\vspace{0.2cm}

\vspace{0.5cm}
\end{center}

\vspace{1cm}


\begin{abstract}

In the lights of current BICEP2 observations accompanied with the PLANCK satellite results,
it has been observed that the simple single field chaotic inflationary models provide
a good agreement with their spectral index $n_s$ and large tensor-to-scalar ratio $r$
($0.15 <r <0.26$). To explore the other simple models, we consider the fractional-chaotic inflationary potentials of the form $V_0 \, \phi^{a/b}$ where $a$ and $b$ are relatively prime.
We show that such kind of inflaton potentials can be realized elegantly in the supergravity framework
with generalized shift symmetry  and a nature bound $a/b < 4$ for consistency. Especially, for the number of e-folding
from 50 to 60 and some $a/b$ from 2 to 3, our predictions are nicely
within at least $1\sigma$  region in the ($r-n_s$) plane.
We also present a systematic investigation of such chaotic inflationary models
with fractional exponents to explore the possibilities for the enhancement
in the magnitude of running of spectral index ($\alpha_{n_s}$) beyond the simplistic models.

\end{abstract}

\clearpage



\section{Introduction}
\label{sec_Intro}
Among the plethora of inflationary models developed so far, the polynomial inflationary potentials have
always been among the center of attraction since the very first proposal as chaotic inflation
in Ref.~\cite{Linde:1983gd}. The recent BICEP2 observations~\cite{Ade:2014xna} interpreted as
the discovery of inflationary gravitational waves have not only taken these models in the limelight
 but also supported these to be the better ones among many others.
The BICEP2 observations fix the inflationary scale by ensuring a large tensor-to-scalar ratio $r$
as follows~\cite{Ade:2014xna}:
\bea
& &  \hskip0.5cm r = 0.20^{+0.07}_{-0.05} ~(68\% ~{\rm CL}) \nonumber\\
& & H_{\rm inf} \simeq 1.2 \times 10^{14} \, \left(\frac{r}{0.16}\right)^{1/2} \, \, {\rm GeV} \, ,
\eea
where $H_{\rm inf}$ denotes the Hubble parameter during the inflation. Subtracting the various dust models and re-deriving the $r$ constraint still results in high significance of detection and one has $ r=0.16^{+0.06}_{-0.05} $. Thus, it suggests the inflationary process to be (a high scale process) near the scale of the Grand Unified Theory (GUT)
 and then can provide invaluable pieces of information on the UV completion proposal such as string theory,
for example, in searching for a consistent supersymmetry (SUSY) breaking scale \cite{Ibanez:2014zsa, Ibanez:2014kia, Harigaya:2014pqa,Biswas:2014kva,Choudhury:2013jya}. On these lines,
some recent progresses on realizing chaotic as well as natural or axion-like inflationary models from string or supergravity framework have been made in Refs.~\cite{Silverstein:2008sg,McAllister:2008hb, Ellis:2014rxa, Hebecker:2014eua, Palti:2014kza, Marchesano:2014mla, Czerny:2014qqa, Kaloper:2014zba, Blumenhagen:2014gta,  Choi:2014uaa, Grimm:2014vva,Ashoorioon:2009wa,Ashoorioon:2011ki,Cicoli:2014sva,Kaloper:2008fb,Kaloper:2011jz}.
However, most of these works on natural as well as chaotic inflation can only produce integral power of
inflaton in the polynomial potential.
In this work, we will present a general fractional chaotic inflation
which can be naturally generated from supergravity framework by utilizing the generalized shift symmetry.

For a given single field potential $V(\phi)$, the sufficient condition for ensuring the slow-roll inflation is encoded in a set of so-called slow-roll conditions defined as follows
\bea
\label{eq:slow}
& & \epsilon \equiv \frac{1}{2} \, \left(\frac{V^\prime}{V}\right)^2 \ll 1 \, , \, \, \eta \equiv \frac{V^{\prime \prime}}{V} \ll 1 \, , \, \,  \xi \equiv \frac{V^\prime \, V^{\prime \prime\prime}}{V^2} \ll 1 \, ,
\eea
where $\prime$ denotes the derivative of the potential w.r.t. the inflaton field $\phi$. Also, the above expression are defined in the units of reduced Planck mass  $M_{\rm Pl} = 2.44 \times 10^{18} \, {\rm GeV}$.

The various cosmological observables such as the number of e-foldings $N_e$, scalar power spectrum $P_s$,
tensorial power spectrum $P_t$, tensor-to-scalar ratio $r$, scalar spectral index $n_s$,
and runnings of spectral index $\alpha_{n_s}$  can be written in terms of the various derivatives of
the inflationary potential via  the slow-roll parameters as introduced above \cite{Lidsey:1995np, Sasaki:1995aw, Gao:2014fva}. For example, the number of e-folding is given as,
\bea
\label{eq:Nefold}
& & N_e \equiv \int_{\phi_{end}}^{\phi_*} \, \frac{1}{\sqrt{2 \epsilon}} \, d \phi \, ,
\eea
where $\phi_{end}$ is determined by the end of the inflationary process when $\epsilon =1$ or $\eta=1$. The other cosmological observables relevant for the present study are given by \cite{Lidsey:1995np, Sasaki:1995aw, Gao:2014fva},
\bea
& & P_s \equiv   \left[\frac{H^2}{4 \, \pi^2 \, (2 \, \epsilon)} \, \left(1 - \left(2 \, C_E - \frac{1}{6}\right) \, \epsilon+\left(C_E \, -\frac{1}{3} \right) \, \eta \right)^2\right]  \, ,\nonumber\\
& & r \equiv  \frac{P_t}{P_s} \simeq 16 \, \epsilon \left[1 -\frac{4}{3} \, \epsilon +\frac{2}{3} \, \eta + 2 \, C_E \, (2 \, \epsilon -\eta)\right] \, , \\
& & n_s \equiv \frac{d \ln P_s}{d \ln k} \simeq 1 + 2 \biggl[\eta - 3 \, \epsilon -\left(\frac{5}{3} + 12 \, C_E \right)\, \epsilon^2 + (8 \, C_E -1) \epsilon \, \eta \nonumber\\
& & \hskip3.5cm + \frac{1}{3}\,\eta^2 -\left(C_E - \frac{1}{3} \right) \, \xi \biggr] \, ,\nonumber\\
& & \alpha_{n_s}\equiv \frac{d n_s}{d \ln k} \simeq 16 \, \epsilon \, \eta - 24 \, \epsilon^2 - 2 \, \xi \, , \nonumber
\eea
where $C_E = -2 + 2 \ln 2 +\gamma = -0.73$, $\gamma =0.57721$ being the Euler-Mascheroni constant. Therefore, it is natural to expect that the shape of the inflationary potential are tightly constrained by the experimental bounds on these cosmological observables coming from various experiments. These experimental constraints on the cosmological parameters are also useful for directly reconstructing the single field potential \cite{Choudhury:2014kma} by fixing the magnitude as well as the various derivatives of the potential. The experimental bounds on cosmological observables relevant in the present study are briefly summarized as follows,
\bea
\label{eq:constrain}
& & r = 0.16^{+0.06}_{-0.05} \, , \, \,  n_s = 0.957 \pm 0.015 \, , \, \, \alpha_{n_s} = -0.022^{+0.020}_{-0.021}~.~
\eea

Apart from the tensor-to-scalar ratio $r$ and spectral index $n_s$, the running of spectral index $\alpha_{n_s}$ has emerged as another crucial cosmological parameter on the lines of reconciling the results between the BICEP2 \cite{Ade:2014xna} and Planck satellite experiments \cite{Ade:2013zuv}. To be more precise, without considering the running of spectral index, Planck + WMAP + highL data \cite{Ade:2013zuv, Hinshaw:2012aka} result in $n_s = 0.9600 \pm 0.0072$ and $r_{0.002} < 0.0457$ at 68 \% CL for the $\Lambda$CDM model, and hence facing a direct incompatibility with recent BICEP2 results. However, with the inclusion of the running of spectral index, the Planck + WMAP + highL data result in $n_s = 0.957 \pm 0.015$, $\alpha_{n_s} = -0.022^{+0.020}_{-0.021}$ and $r_{0.002} < 0.263$ at 95 \% CL. The reconciliation of BICEP2 data along with those of Planck + WMAP + highL \cite{Ade:2013zuv, Hinshaw:2012aka} demands a non-trivial running of spectral index, $\alpha_{n_s} < -0.002$. Although the simplistic single field models are good enough to reproduce the desired values of tensor-to-scalar ratio within $0.15< r < 0.26$ along with $50-60$ e-foldings, most of them fail to generate
large enough magnitude for $\alpha_{n_s}$ (which is needed to be of order $10^{-2}$). On these lines, a confrontation, of realizing the desired values of $n_s, \, r$ and $\alpha_{n_s}$ within a set of reconciled experimental bounds of the BICEP2 and Planck
experiments, has been observed for chaotic and natural inflation models \cite{Gong:2014cqa}.

In this letter, our aim is to consider   the fractional
chaotic inflation models with potentials $V_0 \, \phi^{a/b}$ where $a$ and $b$ are relatively prime.
We shall obtain such kind of inflaton potentials elegantly in the supergravity framework
with generalized shift symmetry for $a/b < 4$ where the shift symmetry is broken
only by superpotential\footnote{Some fractional-chaotic inflationary potentials with a particular value of the exponent can also be derived from supergravity compactifications on 6-dimensional twisted tori \cite{Gur-Ari:2013sba}.}. We find that for the number of e-folding
from 50 to 60 and some $a/b$ from 2 to 3, our models are nicely
within the $1\sigma$ region in the $r-n_s$ plane. Furthermore,
we investigate the possibility of improvements for a better fit of
the three confronting parameters $n_s, \, r$ and $\alpha_{n_s}$ in the lights of Planck and BICEP2 data. However, if the BICEP2
bounds on $r$ is modified/diluted, and if $r$ is found to be smaller than 0.11 in the upcoming
Planck results, then the fractional chaotic inflation can still explain the
data for exponents $a/b < 2$ with $n_s$ being a little bit larger than 0.96.
On the other hand, if the BICEP2 claims would be confirmed in near future,
the open question of interest, for the current model we propose, will be whether one can measure the exponent $a/b$. 

\section{Embedding the Fractional Chaotic Inflation into ${\cal N}=1$ supergravity}
\label{sec_Setup}
In this section, we will provide a supergravity origin of the fractional-chaotic inflationary potential.
The scalar potential in the supergravity theory with given
K\"ahler potential $K$ and superpotential $W$ is
\begin{equation}
V=e^K\left((K^{-1})^{i}_{\bar{j}}D_i W D^{\bar{j}} \overline{W}-3|W|^2 \right)~,~
\label{sgp}
\end{equation}
where $(K^{-1})^{i}_{\bar{j}}$ is the inverse of the K\"ahler metric
$K_{i}^{\bar{j}}=\partial^2 K/\partial \Phi^i\partial{\bar{\Phi}}_{\bar{j}}$, and $D_iW=W_i+K_iW$.
Further the kinetic term for the scalar field is given by
\begin{equation}
{\cal L} ~=~ K_{i}^{\bar{j}} \partial_{\mu} \Phi^i \partial^{\mu} {\bar \Phi}_{\bar{j}}~.~\,
\end{equation}
To warm up, we consider a simple inflation model first in the ${\cal N}=1$ supergravity theory,
where the  K\"ahler potential and superpotential are as follows
\begin{eqnarray}
K=-\frac{1}{2}(\Phi+{\bar \Phi})^2+X{\bar X}-\delta(X{\bar X})^2~,~
\label{KP-A}
\end{eqnarray}
and
\begin{eqnarray}
W~=~Xf(\Phi)~.~\,
\label{SP-A}
\end{eqnarray}
Thus, the K\"ahler potential $K$ is invariant under the shift symmetry~\cite{Kawasaki:2000yn, Yamaguchi:2000vm,
Yamaguchi:2001pw, Kawasaki:2001as, Kallosh:2010ug, Kallosh:2010xz, Nakayama:2013jka, Nakayama:2013txa,
Takahashi:2013cxa, Li:2013nfa}
\begin{eqnarray}
\Phi\rightarrow\Phi+iC~,~\,
\label{SSymmetry-A}
\end{eqnarray}
with $C$ being a dimensionless real parameter, {\it i.e.},
the K\"ahler potential $K$ is a function of $\Phi+\Phi^{\dagger}$, and so it is independent of the imaginary part
of $\Phi$.

The scalar potential can be easily computed from the ansatz of K\"ahler potential $K$ and the superpotential $W$ using Eq.(\ref{sgp})
\begin{eqnarray}
V&=& e^K\left[ |(\Phi + {\bar \Phi})Xf(\Phi)+X \frac{\partial f(\Phi)}{\partial \Phi}|^2+|({\bar X}-2\delta X{\bar X}^2)Xf(\Phi)+f(\Phi)|^2
\right. \nonumber \\ && \left.
-3|Xf(\Phi)|^2\right]~.~\,
\end{eqnarray}
There is no imaginary component ${\rm Im} [\Phi]$ of $\Phi$ in the K\"ahler potential due to
the shift symmetry. Thus, the potential along ${\rm Im} [\Phi]$ is so flat
that it is a natural inflaton candidate.
From the previous studies~\cite{Kallosh:2010ug, Kallosh:2010xz, Li:2013nfa},
we know that the real component
${\rm Re} [\Phi]$ of $\Phi$ and $X$ can be
stabilized at the origin during inflation, {\it i.e.}, ${\rm Re} [\Phi]=0$ and $X=0$.
Therefore, with ${\rm Im} [\Phi]=\phi/{\sqrt 2}$, we obtain the inflaton potential
\begin{eqnarray}
V~=~|f(\phi/{\sqrt 2})|^2~.~\,
\end{eqnarray}
Because superpotential is a holomorphic function of $\Phi$ and $X$, if we choose
\begin{eqnarray}
W~=~ \beta X \Phi^{m/n}~,~
\label{SP-FCI}
\end{eqnarray}
we get
\begin{eqnarray}
V~=~\frac{\beta^2}{2^{m/n}} |\phi|^{2m/n}~.~\,
\end{eqnarray}
One might wonder whether we can select the superpotential in Eq.~(\ref{SP-FCI}) due to
the rational power of $\Phi$.
From a pure supersymmetric theory point of view, it is fine.
In fact, we can derive the K\"ahler potential in Eq.~(\ref{KP-A})
and superpotential in Eq.~(\ref{SP-A}) or Eq.~(\ref{SP-FCI}) from the
K\"ahler potential and superpotential with positive integer powers of all the fields.

Let us consider a superfield $\Phi'$ with the following generalized shify
symmetry~\cite{Takahashi:2010ky, Nakayama:2010kt}
\begin{eqnarray}
\Phi^{\prime n} \rightarrow \Phi^{\prime n}+iC~.~\,
\label{SSymmetry-B}
\end{eqnarray}
The K\"ahler potential and superpotential are
\begin{eqnarray}
K=-\frac{1}{2}(\Phi^{\prime n}+{\bar \Phi}^{\prime n})^2+X{\bar X}-\delta(X{\bar X})^2~,~
\label{KP-B}
\end{eqnarray}
\begin{eqnarray}
W~=~Xf'(\Phi')~.~\,
\label{SP-B}
\end{eqnarray}
And then the kinetic term of $\Phi$ is
\begin{eqnarray}
{\cal L} &=& n^2 (\Phi' {\bar \Phi}')^{n-1} \partial_{\mu} \Phi' \partial^{\mu} {\bar \Phi}'~.~\,
\end{eqnarray}
Unlike the Refs.~\cite{Takahashi:2010ky, Nakayama:2010kt},
we want to emphasize that there is no singularity for the kinetic term of
$\Phi'$ since it is zero if and only if both real and imaginary components of $\Phi'$ vanish.
Thus, adding the extra K\"ahler potential terms of $\Phi$,
as Refs.~\cite{Takahashi:2010ky, Nakayama:2010kt} did, is not necessary at all.
In particular, we obtain the canonical normalized field $\Phi$
\begin{eqnarray}
\Phi \equiv \Phi^{\prime n}~.~\,
\end{eqnarray}
Therefore, the K\"ahler potential in Eq.~(\ref{KP-B}) is the same as that in
Eq.~(\ref{KP-A}), the shift symmetry in Eq.~(\ref{SSymmetry-B}) is the same
as that in Eq.~(\ref{SSymmetry-A}),
and the superpotential becomes
\begin{eqnarray}
W~=~Xf'(\Phi') \equiv X f'(\Phi^{1/n})~.~\,
\end{eqnarray}

\begin{table}[t]
\begin{center}
\begin{tabular}{|c|c|c|}
  \hline
 & ~ $\Phi'$ ~& ~$X$~ \\
    \hline
~$U(1)_R$~ & $0$ & $2$    \\
   \hline
~$Z_{2n}$~ & $1$ & $-m$   \\
   \hline
   \end{tabular}
\end{center}
\caption{The quantum numbers of $\Phi'$ and $X$ under the $U(1)_R$ and $Z_{2n}$ symmetries.
Here, $m < 2n$ is required.}
\label{QN-XPhiP}
\end{table}

Now, let us give the concrete models, which can realize the superpotential
in Eq.~(\ref{SP-FCI}). We consider $U(1)_R$ symmetry and introduce the $Z_{2n}$ symmetry.
The respective quantum numbers for $\Phi'$ and $X$ are given in Table~\ref{QN-XPhiP}.
Especially, we want to emphasize  $0 < m < 2n$ for consistency. One can easily show
 that the K\"ahler potential, which is consistent with the $U(1)_R$ and $Z_{2n}$ symmetries,
is given by Eq.~(\ref{KP-B}) up to the higher order terms, and the superpotential is
\begin{eqnarray}
W~=~\beta X \Phi^{\prime m}~.~
\end{eqnarray}
With the canonical normalization of $\Phi'$, we indeed obtain the superpotential in Eq.~(\ref{SP-FCI}).
In addition, for the inflation potential $V_0 \, \phi^{a/b}$ which will be studied
in the following, we get a nature bound as
\begin{eqnarray}
\frac{a}{b} & \equiv & \frac{2m}{n} ~<~ \frac{4n}{n}~=~4~.~
\end{eqnarray}
To fit the Planck and BICEP2 data, we need $a/b$ from 2 to 3 on which we will elaborate later on while discussing the numerical results for the various cosmological observables. Thus, this consistency condition ($a/b <4$) from
supergravity model building can be obviously satisfied.
However, if the BICEP2 claims about $r$ is modified (or if the bounds on $r$ are diluted), and $r$ is found to be smaller than 0.11 from
the upcoming Planck results, then our fractional chaotic inflation can explain the
data only for $a/b < 2$ and $n_s$ being a little bit larger than 0.96.

\section{Fractional Chaotic Inflation: Numerical Study}
 From last section, we end up with a single field inflationary potential of chaotic-inflation type monomials with fractional exponents as
\bea
\label{eq:Vpot}
& & V(\phi) = V_0 \, \phi^{a/b}~,~
\eea
where $a$ and $b$ are positive integers and $a/b < 4$. In the subsequent analysis, our main focus and the special attention would be to study the inflationary models in which $a$ and $b$ are coprime. The sufficient condition for the ensuring
the inflationary process is encoded in terms of so-called slow-roll conditions
\bea
\epsilon \ll 1, \, \eta \ll 1 , \, \, \, \, {\rm and} \,\,\,\,\, \, \xi \ll 1. \nonumber
\eea
Now the various slow-roll parameters (as defined in Eq.(\ref{eq:slow})) simplify for the potential (\ref{eq:Vpot}) as follows
\bea
\epsilon  = \frac{a^2}{2 \, b^2 \, \phi^2}, \,\,
 \eta  = \frac{a \, (a-b)}{b^2 \, \phi^2}\, ,\,\,
 \xi = \frac{a^2 \, (a-b)\, (a-2b)}{b^4 \, \phi^4}~.~\,
\eea
The number of e-folding generated between the phase of horizon exit and the end of inflation is calculated by Eq.(\ref{eq:Nefold}) as
$N_e  = \frac{b}{2 \, a} \, \left(\phi_*^2 - \phi_{end}^2\right)~,~
$
with $\phi_{end} = \frac{a}{\sqrt2 \, b}$. 
The simplified expressions of the cosmological observables $n_s, r$ and $\alpha_{n_s}$ for the present class of models are
\bea
& & r = \frac{8 \, a^2 \left[a \, (6 \, {C_E}-2) + \, 3 \, b \, \phi^2\right]}{3 \, b^3 \, \phi^4} \, , \nonumber\\
& & n_s =\frac{1}{6\, b^4 \,\phi^4} \biggl[a^4 \, (12 \,C_E-7)-2 \, a^3 \, b \,(24 \,C_E + 1) \, \\
& & \hskip3cm + \, 2 \, a^2 \, b^2 \, \left(2-3 \, \phi^2\right)-12 \, a \, b^3 \, \phi^2+ \, 6 \, b^4 \, \phi^4 \biggr]\, ,\nonumber\\
& & \alpha_{n_s}=-\frac{2 \, a^2 \, (a + 2 \, b)}{b^3 \, \phi^4} . \nonumber
\eea
From the above expressions, one naively observes the possibility of enhancing the magnitude of the running of spectral index $\alpha_{n_s}$ by a choice of ratio $a/b \gg 1$. However, it is not as arbitrarily possible as it seems to be, since large ratio of $a/b$ will lead the spectral index ($n_s$) values going outside the experimentally allowed window. With introducing a new parameter $p = \frac{b}{a}$, all the cosmological parameters defined earlier can be equivalently written out in terms of the model dependent parameter $p$ and one of the observables in $\{\phi, N_e, r , n_s, \alpha_{n_s}\}$. A few set of expressions are presented as follows.

\subsubsection*{Observables/parameters in terms of $p$ and $\phi$}
\bea
& & \epsilon = \frac{1}{2 \, p^2 \, \phi ^2}, \, \, \, \eta = \frac{1-\, p}{p^2 \, \phi ^2}, \, \, \, \, \xi = \frac{(p-1) \, (2 p-1)}{p^4 \, \phi ^4}, \, \, \, \,\\
& &  N_e =\frac{1}{2} \, p \, \left(\phi ^2- \, \frac{1}{2 \, p^2}\right)\, , \, \, \, \, r =\frac{8 \left(6 \, {C_E}+ 3 \, p \, \phi^2 \, - 2 \right)}{3 \, p^3 \, \phi ^4} \, ,\nonumber
\eea
\bea
& & n_s =\frac{{C_E} (12 - 48 \, p)+6\, p^4 \, \phi^4 - 12 \, p^3 \, \phi^2 + p^2 \, \left(4 - 6 \, \phi^2 \right) - 2\,  p- 7}{6\, p^4 \,\phi^4}, \nonumber\\
& & \alpha_{n_s} = -\frac{4 \, p + 2}{p^3 \, \phi^4}. \nonumber
\eea
\subsubsection*{Observables/parameters in terms of $p$ and $N_e$}
\bea
\label{eqsNe}
& & \epsilon = \frac{1}{1 + 4 \, p \, N_e}, \, \, \, \eta = \frac{2 - 2 \, p}{1 + 4 \, p \, N_e}, \, \, \xi = \frac{4 \, (p -1) \, (2 \, p-1)}{\left(4 \, p \, N_e + 1 \right)^2}\, \, ,\\
& & \phi = \frac{\sqrt{1 + 4 \, p \, N_e}}{p \, \sqrt2}, \, \, \,  r = \frac{16 \, \left(4 \, (3 \, {C_E} - \,1) \, p + 12\, p \, N_e + 3\right)}{3 \, \left(4 \, p \, N_e + 1\right)^2} \, ,\nonumber\\
& & n_s =\frac{\left(24 \, C_E -17\right) -16\,p \left(1 + 6 \, C_E\right) \,+ 8 \, p^2 \bigl[1 -6 N_e + 6 \, N_e^2 \bigr]}{3 \, (1 + 4 \, p \, N_e)^2} , \nonumber\\
& & \alpha_{n_s} = -\frac{8 \, p \, (1 + 2 \, p)}{(1 + 4 \, p \, N_e)^2} .\nonumber
\eea
\subsubsection*{Observables/parameters in terms of $p$ and $r$}
\bea
& & \epsilon = \frac{3 \, r}{8 \, (Y + 3)}, \, \, \, \eta = -\frac{3 \, (p - 1)\, r}{4 \, (Y + 3)} \, , \, \, \xi = \frac{9\, (p-1) \, (2 \, p-1) \, r^2}{16 \, (Y + 3)^2}\, \, ,\\
& & \phi^2 = \frac{12 \pm 4 \, Y}{3 \, p^2 \, r}, \, \, \, N_e = \frac{-3 \, r+ 8 \, Y+ \, 24}{12 \, p \, r} \, ,  \nonumber\\
& & n_s =\frac{1}{32 \, (Y + 3)^2} \, \biggl[3 \, \biggl\{12 \, C_E \,  r  \, (r - 4 \, p \, (r-2)) + \, \left(4 \, p^2 - 2 \, p - 7 \right) \, r^2 \nonumber\\
& & - 8 \, r \, (2 \, p \, (Y + 5) + Y + 3) + 64 \, (Y + 3)\biggr\}\biggr] , \, \, \, \alpha_{n_s} = -\frac{9 p (2 p+1) r^2}{8 (Y+3)^2} \,\nonumber
\eea
where $Y = \pm \sqrt{9 - 3 p\, r\, + 9 \, p \, r \, C_E}$. Imposing $Y$ to be a real number will result in the following inequality
\bea
p<\frac{0.94}{r}, \, \, \, \hskip1cm i.e. \hskip0.8cm \frac{a}{b} > \frac{r}{0.94}~,~
\eea
and hence gives an important bound on choice of fractions. Similarly, the expressions for other combinations $\{p, \, n_s \}$ and $\{p , \alpha_{n_s}\}$ can also be written out. Now, looking for the solutions under the constraint Eq.(\ref{eq:constrain}), we find the constraint on $\alpha_{n_s}$ to be the hardest one to match. This is also consistent with the previous study in \cite{Gong:2014cqa}. However, we manage to get $\alpha_{n_s} \sim - 0.00108$ for $N_e =60 $ with $a/b \leq 3$, which is relatively better than the standard single field inflationary model. To illustrate this numerically, let us consider the set of Eq. (\ref{eqsNe}) and fix the number of e-foldings to sixty, then solving the following inequalities sequentially, we get
\bea
0.953 < n_s < 0.967 \, \, \& \, \, p >0 \Longrightarrow 0.276833 < p < 0.499343 \nonumber\\
 0.15 < r < 0.26 \, \, \& \, \, p >0 \Longrightarrow 0.247775 < p < 0.432476  \\
\alpha_{n_s} < -0.00108  \, \, \& \, \, p >0 \Longrightarrow 0.000142779 < p < 0.278093. \nonumber
\eea
Thus, we observe that one can easily have $n_s \simeq 0.96, r \simeq 0.2$ and $N_e \simeq 60$, but with $|\alpha_{n_s}| <  0.001$. There does not exist a solution for $p$ if one considers larger value for $|\alpha_{n_s}|$ in the third inequality. In order to have larger magnitude of  $\alpha_{n_s}$, one has to compromise with the number of e-folding which is entangled with the spectral index $n_s$. For a given $a/b$, the value of $n_s$ is lowered by lowering the number of e-foldings, and it will be outside the experimental bounds for a certain value of $N_e$.
 The ratio $a/b$ along with any one of $\{ N_e, \phi, r , n_s, \alpha_{n_s} \}$ generically fixes all the rest  observables in this single field setup.
 Some of the possibilities are tabulated in Table~\ref{tab1}.  These numerical observations will be clearer in the graphical analysis as we discuss now.
\begin{table}[h!]
  \centering
  \begin{tabular}{|c|c||c|c|c||c|c|c|}
  \hline
    $N_e$   & $a/b$ & $\epsilon$   & $\eta$ & $\xi$ &$n_s$  & $r$  & $\alpha_{n_s}$   \\
    \hline \hline
    &&&&&&&\\
     60 & $\frac{10}{3}$ &0.014&0.019&0.0002& 0.956 & 0.216 & -0.00072  \\
     \hline
    &&& &&&&\\
    50 & $\frac{5}{2}$ &0.012&0.015&0.00007& 0.955 & 0.194 & -0.00088  \\
   \hline
   &&&&&&&\\
    40 & $\frac{5}{3}$&0.010&0.008&-0.00003 & 0.955 & 0.161 & -0.00112  \\
    \hline
  \end{tabular}
  \caption{Some sampling values for cosmological observables and parameters.}
  \label{tab1}
 \end{table}

\begin{figure}[H]
\begin{center}
\includegraphics[width=9cm]{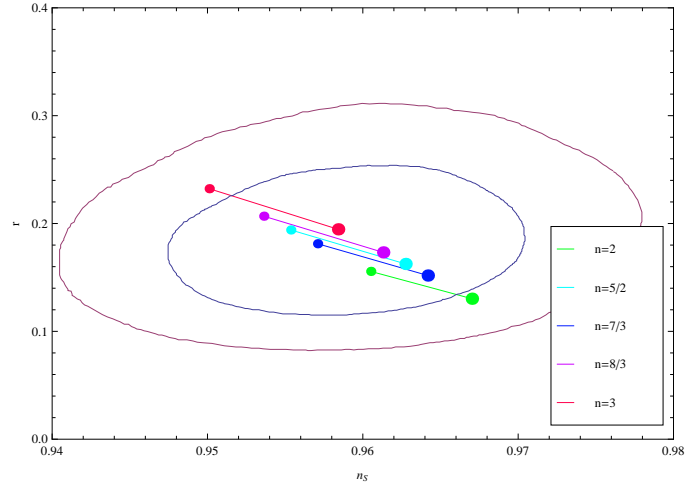}
\caption{ The $n_s-r$ plot for fractional chaotic potentials with $n \equiv a/b= 2, 5/2, 7/3, 8/3$ and 3.
The blue and red circles are for the 68$\%$   and the 95$\%$ CL regions for $r$ and $n_s$. The number of the e-folding is from 50 to 60 .}
\label{Fig:ns-r}
\end{center}
\end{figure}

\begin{figure}[H]
\begin{center}
\includegraphics[width=9cm]{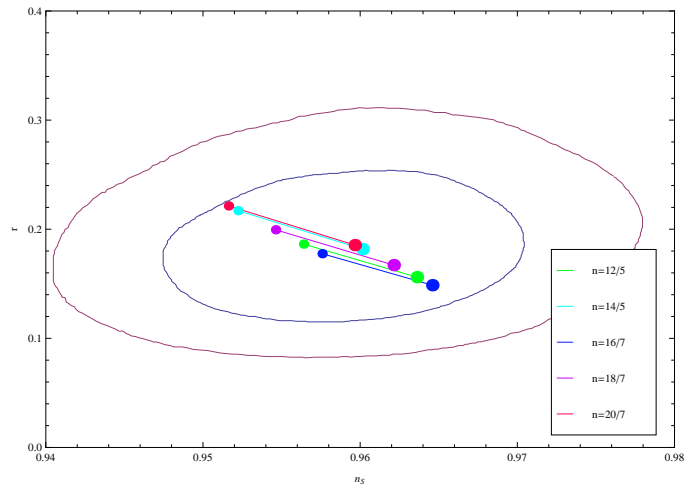}
\caption{ The $n_s-r$ plot for fractional chaotic potentials with $n \equiv a/b= 12/5, 14/5, 16/7, 18/7$ and 20/7.
The blue and red circles are for the 95$\%$   and the 68$\%$ CL regions for $r$ and $n_s$. The number of the e-folding is from 50 to 60 .}
\label{Fig:ns-r-N}
\end{center}
\end{figure}

\begin{figure}[H]
\begin{center}
\includegraphics[width=9cm]{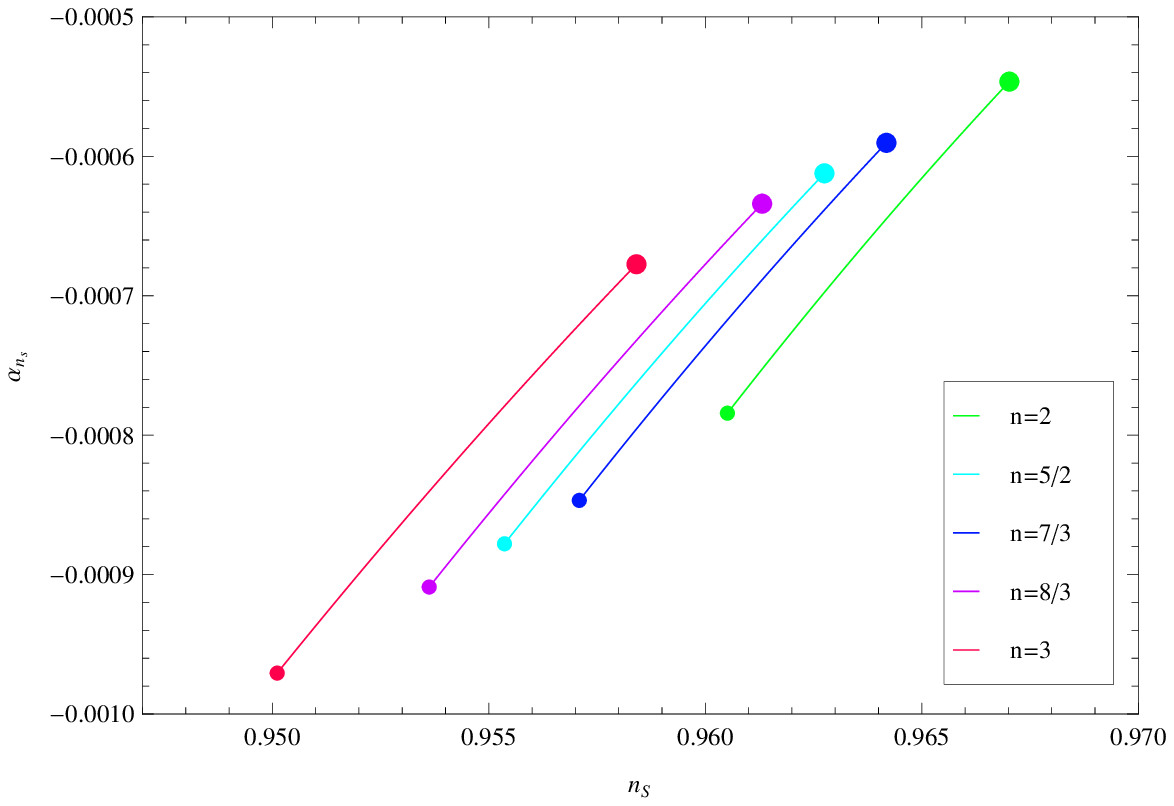}
\caption{ The $n_s$-$\alpha_{n_s}$ plot for fractional chaotic potentials
with $n \equiv a/b= 2, 5/2, 7/3, 8/3$ and 3.
 The number of e-folding is from 50 to 60 .  }
\label{Fig:ns-dns}
\end{center}
\end{figure}

\begin{figure}[H]
\begin{center}
\includegraphics[width=9cm]{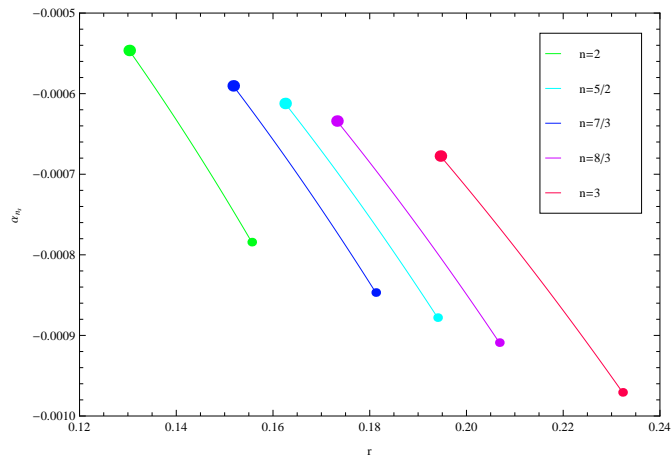}
\caption{ The $r-\alpha_{n_s}$ plot  for fractional chaotic potentials
with $n \equiv a/b= 2, 5/2, 7/3, 8/3$ and 3. The colors are the same
as these in Fig.~\ref{Fig:ns-dns}. The number of e-folding is from 50 to 60 .  }
\label{Fig:r-dns}
\end{center}
\end{figure}

The Figs.~\ref{Fig:ns-r}
and \ref{Fig:ns-r-N}  represent the relations ($n_s$ - $r$) indicating that it is fairly possible to have well consistent $n_s$ and $r$ values for chaotic inflationary models including integer as well as fractional exponents. In particular, for $a/b=5/2, 7/3, 8/3, 12/5, 14/5, 16/7$ and 18/7, our predictions are within
the $1\sigma$ region for the number of e-folding in the range from 50 to 60. However, as we have argued earlier that reconciling the Planck data with the recent BICEP2 observations demands a non-trivial running ($\alpha_{n_s}$) of the spectral index ($n_s$). It is in confrontation with desired $n_s$ and $r$ values as can be well seen from the {\it opposite slopes} of lines plotted for different exponents in Fig.~\ref{Fig:ns-dns} and Fig.~\ref{Fig:r-dns}. Actually, this confrontation has been recently studied in detail in \cite{Gong:2014cqa} in the context of polynomial-chaotic inflation \cite{Nakayama:2013txa, Linde:1981mu}, Natural inflation models \cite{Freese:1990rb} as well as S-dual inflation models \cite{Anchordoqui:2014uua}.

\begin{figure}[H]
\begin{center}
\includegraphics[width=9.cm]{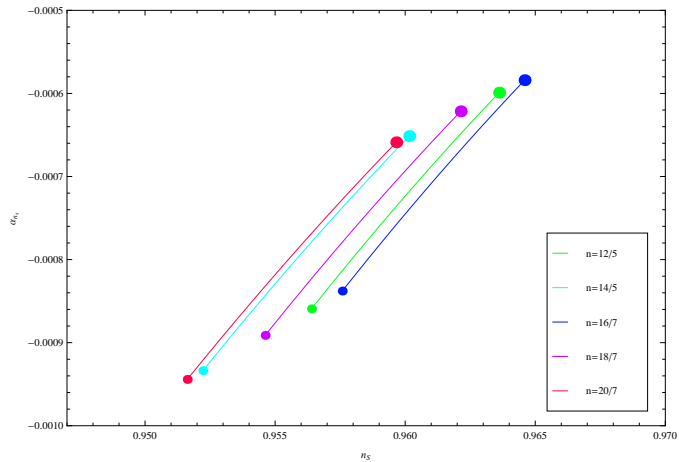}
\caption{ The $n_s$-$\alpha_{n_s}$ plot for fractional chaotic potentials
with $n \equiv a/b= 12/5, 14/5, 16/7, 18/7$ and $20/7$.
 The number of e-folding is from 50 to 60 .  }
\label{Fig:ns-dns2}
\end{center}
\end{figure}

The Fig.~\ref{Fig:ns-dns} and Fig.~\ref{Fig:r-dns} in our analysis recover that for standard $\phi^2$ chaotic inflation \cite{Linde:1981mu}, $\alpha_{n_s} = -8 \times 10^{-4}$ is the best value corresponding to $N_e = 50$. However, we can see that a larger value of $a/b$ will result in a better situation for relaxing the tension between BICEP2 and PLANCK data.  The magnitude of running of spectral index $\alpha_{n_s}$ increases together with the exponents up to $\alpha_{n_s} = - 0.001$ for $N_e = 50$ with a $\phi^3$ potential, but it takes the spectral index ($n_s$) values into the marginal regime. Also, the ($\alpha_{n_s}-n_s$)  relation for some more fractional exponents with intermediate values $2<a/b<3$ are plotted in Fig[\ref{Fig:ns-dns2}].

From the plots in Fig.~\ref{Fig:ns-dns} -Fig.~\ref{Fig:ns-dns2}, we can see that the best fit with the BICEP2 experiment is with fractional powers such as $8/3$ or $20/7$ lying between 2 and 3, which is giving a reasonable (however not very large) enhancement in the negative running of spectral index $\alpha_{n_s}$.

\section{Conclusions}
\label{sec_Conclusions}

In this article, we have constructed the fractional-chaotic inflationary potentials from
the supergravity framework utilizing the generalized shift symmetry. A nature bound on the fractional power is provided in our model for consistency.  One of the motivations for this construction is to investigate the possibility of realizing a non-trivial and negative running ($\alpha_{n_s}$) of the spectral index $n_s$.
It has been well established by now that the simplistic polynomial chaotic inflation models successfully realize large tensor-to-scalar ratio $r$
compatible with the allowed values of the spectral index $n_s$. 
However, in order to reconcile the data from the Planck and recent BICEP2 observations,
a non-trivial running of spectral index is needed which is usually suppressed
at order $10^{-4}$ in chaotic inflationary models. We have studied a generalization of polynomial chaotic inflation by including the fractional exponents in search for possible improvements along this direction. We found that, for the number of e-folding
from 50 to 60 and some fractional exponents $a/b$ from 2 to 3, our results are nicely
within the $1\sigma$ region in the ($r-n_s$) plane along with an improvement (although insufficient) in the magnitude of the running of spectral index ($\alpha_{n_s}$). Such a class of fractional-chaotic inflationary potentials can also be interesting for facilitating large field excursions on the lines of \cite{Harigaya:2014eta}, and also for studying other cosmological aspects on the lines of \cite{Bastero-Gil:2014oga,Bartrum:2013fia,Bastero-Gil:2014jsa}. If the BICEP2 result will be confirmed in (near) future,
the open question, relevant to our fractional-chaotic inflationary model, will be whether one can measure the exponent $a/b$ at some future
experiments. 

\section*{Acknowledgments}

We would like to thank Yungui Gong and Chi Tian very much for helpful discussions. TL and XG was supported in part by the Natural Science Foundation of China under grant numbers 10821504, 11075194, 11135003, and 11275246, and by the National Basic Research Program of China (973 Program) under grant number 2010CB833000. PS was supported by the Compagnia di San Paolo contract ``Modern Application of String Theory'' (MAST) TO-Call3-2012-0088.


\nocite{*}
\bibliography{ns-r}
\bibliographystyle{utphys}


\end{document}